\documentclass[12pt,preprint]{aastex}

\usepackage{amsmath,amsfonts,graphicx,color}

\slugcomment{}

\shorttitle{}
\shortauthors{Camporeale \& Burgess}

\begin{document}


\title{The dissipation of solar wind turbulent fluctuations at electron scales}


\author{Enrico Camporeale\altaffilmark{1} and David Burgess}
\affil{Queen Mary University of London, Mile End Road, London E1 4NS, United Kingdom}
\altaffiltext{1}{now at the Applied Mathematics and Plasma Physics Group, Theoretical Division, Los Alamos National Laboratory, Los Alamos, NM 87545 }
\begin{abstract}
We present two-dimensional fully-kinetic Particle-in-Cell simulations of decaying electromagnetic fluctuations. The computational box is such that wavelengths ranging from electron to ion gyroradii are resolved. The parameters used are realistic for the solar wind, and the ion to electron mass ratio is physical. The understanding of the dissipation of turbulent fluctuations at small scales is thought to be a crucial mechanism for solar wind acceleration and coronal heating. The computational results suggest that a power law cascade of magnetic fluctuations could be sustained up to scales of the electron Larmor radius and smaller. We analyse the simulation results in the light of the Vlasov linear theory, and we comment on the particle heating. The dispersion curves of lightly damped modes in this regime suggest that a linear mechanism could be responsible for the observed steepening of power spectra at electron scales, but a straightforward identification of turbulent fluctuations as an ensemble of linear modes is not possible.
\end{abstract}


\section{Introduction}

The solar wind is a collisionless medium which is known to be turbulent over many length and time scales. The first observations of an active turbulent cascade at large scales in the solar wind date back to the work of \citet{coleman68}. In analogy with hydrodynamic turbulence, the large scale fluctuations are expected to nonlinearly cascade to smaller scales and being ultimately dissipated in the so-called dissipation range. However, in a collisionless plasma ordinary viscosity cannot play a decisive role for the dissipation of energy, as it does in a neutral fluid. Hence, it has been long thought that the onset of dissipation range in the solar wind must be associated with cyclotron or Landau damping, and therefore must lie within kinetic scales.\\
The understanding of the mechanisms that convert the electromagnetic turbulent energy into kinetic heating is far from complete. This issue is very relevant for the development of a realistic scenario of solar wind acceleration and coronal heating.\\
The non-hydrodynamic nature of the physics involved in the dissipation process is now widely appreciated \citep{smith06a}, and the understanding of dissipation mechanisms at small scales represents a very active field of research.\\
Observations indicate that kinetic plasma physics at both proton and electron scales are involved \citep{matthaeus08}. Moreover, it is known that the Kolmogorov-like inertial range power spectrum terminates at $\sim 0.3$ Hz, where the spectrum of magnetic fluctuations steepens to form a power law with an average index close to $-3$ and varying between $-2$ and $-5$. \citep{leamon98b, leamon98a, leamon99}. More recent observations \citep{sahraoui09,sahraoui10} have shown that the power law in the range of scales between ion and electron gyroradii is quite universal, and close to 2.8. They also suggested the possibility that a nonlinear turbulent cascade could proceed further above the frequency associated with electron gyromotion, where the power spectrum supposedly undergoes another abrupt break and steepening. This interpretation has been challenged by observations reported in \citet{alexandrova09}, that suggest an exponential roll-over of the power spectrum at electron scales, which is usually interpreted as the onset of linear dissipative mechanisms.\\
In the last few years an intensive effort has been directed to understand the mechanism that regulates the damping of turbulent fluctuations at small scales through a number of different approaches involving numerical simulations. In particular, \citet{howes08b} have performed state of the art gyrokinetic simulations of kinetic Alfven waves (KAW) cascade; Particle-In-Cell (PIC) simulations of decaying whistler modes have been performed by \citet{saito08}; \citet{svidzinski09} have presented PIC simulations of fast magnetosonic waves. Hybrid simulations have been presented by \citet{parashar09} and \citet{markovskii10}, and a shell model for 3D Hall-MHD has been studied by \citet{galtier07}.\\
Interestingly, many works in this area have made the assumption that there must be a certain length scale where the amplitude of magnetic fluctuations becomes so small that it is justifiable to treat the turbulent dissipation as damping of linear waves. Therefore the spectrum of magnetic fluctuations has been tentatively compared to linear theory predictions in a number of papers  \citep{li01, stawicki01, howes08a, gary08, jiang09, podesta10}. Here it is (sometimes implicitly) assumed that linear damping mechanisms can convert electromagnetic energy to kinetic energy with the same efficiency and rate in a turbulent system and in an homogeneous plasma in thermodynamical equilibrium. This approach has been recently challenged in \citet{camporeale10}, where it is argued that using the least damped modes of a normal mode linear analysis might be a misleading way to characterise the dissipation.\\
In any case, if any linear mode is dominant in regulating the cascade in the dissipation range, there is certainly no agreement on which mode that should be \citep{gary09}.
It has to be mentioned that the numerical approach mentioned above and pursued by many researchers, has so far focused on ion scales, not extending very deeply into electron scales, mainly for reasons of computational costs.\\
The aim of this paper is to present fully-kinetic PIC simulations of decaying electromagnetic fluctuations at electron scales. In doing so we address the following open problems. It is still unknown whether a linear approximation for the damping rate of the fluctuations, that would effectively describe the waves as an ensemble of linear modes, is fully justifiable and at which scale that approximation becomes valid. Moreover, there is an open debate about the scale at which such fluctuations are expected to be completely dissipated, and what mechanism is responsible for the dissipation. As we will show, our results do not support a scenario where turbulent fluctuations can be described as an ensemble of the least damped linear waves, not even at very small scales, although the observed steepening of the power spectra might be associated with the steepening of the dispersion curves of linear modes.\\
The feasibility of the simulations presented in this paper has been greatly enhanced by using a semi-implicit method both for the field solver and the particle mover, and we will discuss some computational issues that have emerged in this study, and that should be taken into account for future studies.\\
The paper is organised as follows. Section 2 briefly describes the PIC code, the parameters used, the issues related to the initialization of the simulations, and compares this paper with previous works. The main results are described in Section 3. The emphasis will be on the observed nonlinear cascade, the identification of linear modes, and particle heating. The conclusions and discussion of future work are given in Section 4.

\section{Methodology}
\begin{figure}
 \center\includegraphics[width=25pc]{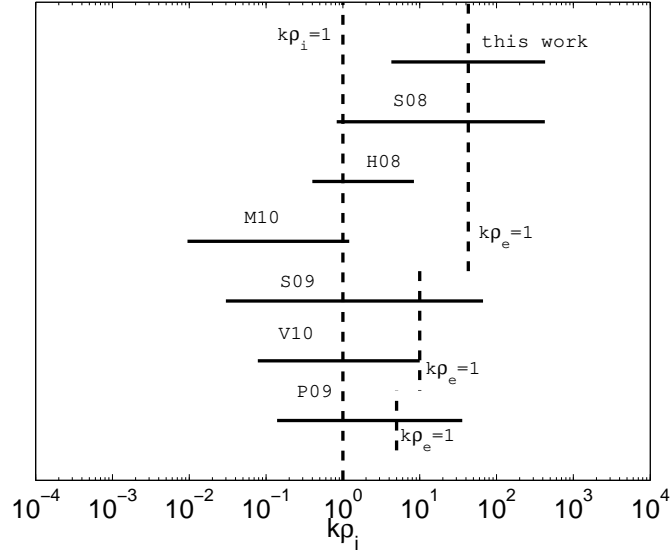}
 \caption{Comparison of the range of wavevectors studied in different works. The acronym list is reported in Table \ref{tab1}. The vertical dashed lines indicate the inverse proton Larmor radius ($k\rho_i=1$), and the inverse electron Larmor radius ($k\rho_e=1$), assuming equal ion and electron temperatures. Notice that the latter is shifted in works that do not use a physical mass ratio. Also, notice that V10 has reported simulations with different temperature ratios (and therefore different locations of $k\rho_e=1$).}\label{fig0}
 \end{figure}

We perform simulations of an ion-electron plasma in a two-dimensional periodic box in Cartesian geometry $(x,y)$. The code used is a fully-kinetic, electromagnetic, parallel PIC code, called \textit{PARSEK2D}. The main feature of the code is the use of an implicit moment method for advancing the fields in time and a predictor-corrector routine to advance the particles. A thorough description of the algorithm and the code can be found in \citet{markidis09}. The implicitness of the scheme allows a relaxation of the stability conditions typical of an explicit code. In particular the cell size $\Delta x$ is not required to be comparable with the Debye length, the timestep $\Delta t$ can be larger than the inverse plasma frequency and, in general, the factor $c\Delta t/\Delta x$ (where $c$ is the speed of light) can be larger than one.\\
For the simulations presented in this paper, we have found that it is still convenient, for accuracy reasons, to resolve the inverse electron plasma frequency. The cell size $\Delta x$, however, is about 20 times the Debye length, and the Courant factor $c\Delta t/\Delta x$ is about 9. Therefore the total saving with respect to an explicit code is at least a factor of 80,000 (not taking into account the complexity of the algorithm). This large saving factor translates into the possibility of simulating a plasma in realistic conditions in a computational box which is fairly large. We compare in Figure \ref{fig0} the range of wavevectors included in the computational box for the simulations presented in this paper with previous works. Table \ref{tab1} compares other relevant parameters, such as the ratio of plasma to cyclotron frequency (this is undefined in hybrid simulations), and the ion to electron mass ratio. Notice that the use of an artificial mass ratio makes the values of the ion and electron gyroradii closer to each other, resulting in an extension of the range of wavevector studied. This of course results in a computational advantage, but it is often difficult to estimate the effects of this approximation on the results, since it is very well known that the separation of ion and electron scales is an intrinsic and important characteristic of plasmas. What emerges from Figure \ref{fig0} is that, as far as we know, this paper presents the first simulations that use realistic parameters in a fairly large computational domain that extends to electron scales. We point out that also the work by \cite{saito08} was in principle able to extend to electron scales, but the number of particles used was too low in order to distinguish the results above $k\rho_i\sim 10$ from the noise level.\\ 
In this work, the plasma is in solar wind conditions. The electron plasma beta is equal to 0.5 ($\beta_e=8\pi n_0T_e/B_0^2$, $T_e$ is the electron temperature, $n_0$ is the density, and $B_0$ is the magnetic field intensity). Ions and electrons have initially equal temperature, and a Maxwellian distribution function. The ratio of ion plasma frequency to ion gyrofrequency $\omega_{pi}/\Omega_{i}$ is about 1650, and the ion to electron mass ratio is physical ($m_i/m_e=1836$). The number of cells is $200\times 200$, and the box size is large enough to accommodate waves with wavenumbers $k$ ranging from $k\rho_i=4.28$ (equivalent to $k\rho_e=0.1$), to $k\rho_e=10$, where $\rho_i$ and $\rho_e$ are the ion and electron gyroradius, respectively. In all the simulations presented in this paper we have used 6400 particles per cell, for each species, and the simulations are typically run for a time $T=100\Omega_{e}^{-1}$ ($\Omega_{e}$ is the electron gyrofrequency).  \\
The choice of initial conditions offers a wide degree of freedom. The general idea is to initialize a spectrum of fluctuations with properties somehow similar to the one observed in the solar wind, and follow their decay in time. Some authors have performed simulations where a spectrum of only one plasma linear mode was excited at small values of $k$ (for instance, KAW or whistlers). In order to do so, one should compute the phase and amplitude relations between different quantities such as magnetic and electric field, velocity and density perturbation predicted by linear theory for any single value of $k$, and impose such perturbations to the background equilibrium. In order to get a completely self-consistent initial perturbation, the particle distribution function should be perturbed exactly. This is very expensive, from a computational point of view, and a shortcut which is very often employed is to initialize the particles with a shifted Maxwellian distribution, such that only the current and charge densities are consistent with the linear theory predictions, but not all the higher-order moments of the distribution function. The implicit assumption of such a method is that the PIC code will be able to quickly `self-adjust' the initial inconsistency, but this approach remains, in our view, not very satisfactory. Moreover, the choice of the slope of the initial power spectrum constitutes yet another degree of freedom in the initialization.\\
In this work, we have opted for a very general initial perturbation. In this way, we do not face the problem of choosing some particular modes to initialize, and we are also able to study which mode, if any, `survives' through the nonlinear cascade. Accordingly, we initialize our simulations with a background magnetic field in the $x$ direction, and we superimpose a spectrum of random phase fluctuations in the magnetic field only. We assume that such perturbation is general enough to perturb a large variety of waves in the plasma. We point out that, differing from the normal mode initialization described above, this choice is self-consistent both with Vlasov and Maxwell equations. The plasma will certainly respond instantaneously to magnetic field gradients by creating currents and generating electric fluctuations, and ultimately waves. The only requisite that the initial perturbation must satisfy is to be divergence-free. Hence, we initialize the magnetic perturbations as: 
\begin{multline}\label{1}
\delta B_x= \sum_{k_x} \sum_{k_y} |k_x|^{-(\alpha+1)} |k_y|^{-\beta} \cos(k_xx+k_yy+\phi_{x,y}) +  
\sum_{k_y} \left|\frac{2\pi}{L_x}\right|^{-(\alpha+1)} |k_y|^{-\beta} \cos(k_yy+\phi_{x})
\end{multline}
\begin{multline}\label{2}
\delta B_y= -\sum_{k_x} \sum_{k_y} |k_x|^{-\alpha} |k_y|^{-(\beta+1)} \cos(k_xx+k_yy+\phi_{x,y}) +  
\sum_{k_x} |k_x|^{-\alpha}\left|\frac{2\pi}{L_y}\right|^{-(\beta+1)} \cos(k_xx+\phi_{y})
\end{multline}
\begin{equation}\label{3}
\delta B_z= \sum_{k_x} \sum_{k_y} \frac{|k_x|^{-\alpha} |k_y|^{-\beta}}{(k_x+k_y)} \cos(k_xx+k_yy+\phi_z)
\end{equation}
where $L_x$ and $L_y$ are the box length, $k_x=2\pi m/L_x$, $k_y=2\pi n/L_y$, $m$ and $n$ are integers which usually range from $-3$ to $3$, and $\phi_{x,y}, \phi_{x}, \phi_{y}, \phi_z$ are random angles. Note that the choice of $\delta B_z$ is arbitrary and is made in such a way that its amplitude is of the order of the amplitudes of $\delta B_x$ and $\delta B_y$. We have run several simulations with different values for the slopes $\alpha$ and $\beta$ of the spectra and the results seem not to be significantly affected by those parameters, For simplicity, we will show the results when $\alpha=\beta=0$.\\
As a further check on the soundness of the results we have also initialized some simulations with a spectrum of waves with an Alfvenic velocity perturbation imposed, and the results are qualitatively similar, for the length and time scales investigated here.

\section{Results}
\subsection{Turbulent cascade}
\begin{figure}
 \center\includegraphics[width=25pc]{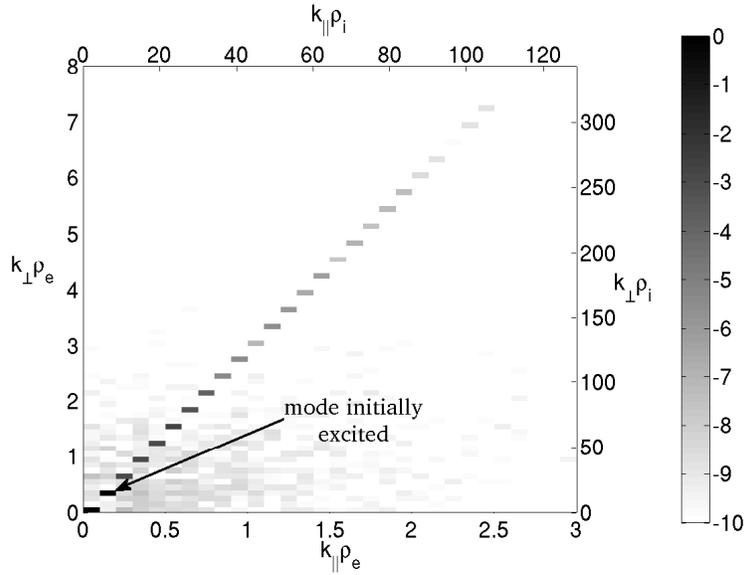}
 \caption{Spectrum of magnetic fluctuations $\delta B/B_0$ in wavenumber space, in logarithmic scale. The mode initially excited $m=1,n=3$ is indicated with an arrow.}\label{fig1}
 \end{figure}

\begin{figure}
 \center\includegraphics[width=25pc]{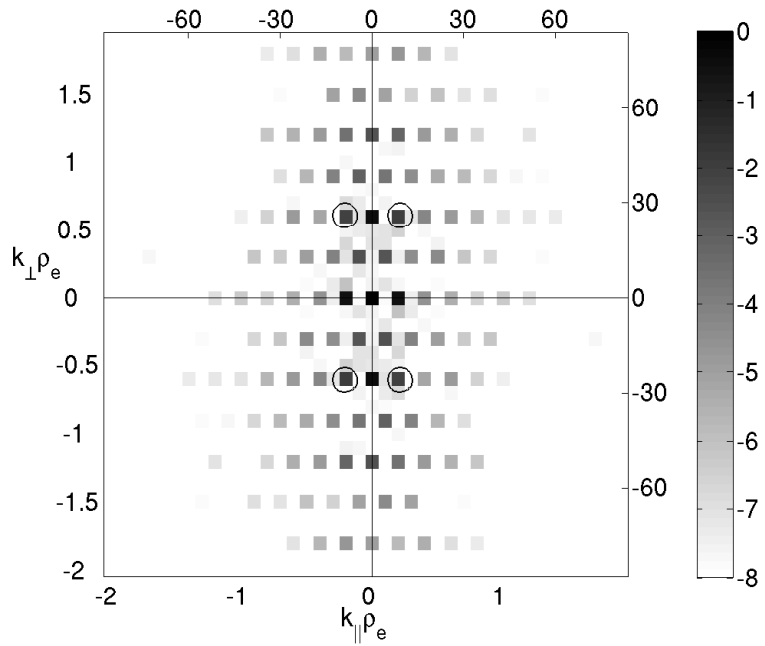}
 \caption{Spectrum of magnetic fluctuations $\delta B/B_0$ in wavenumber space, in logarithmic scale. The modes initially excited $m=1,n=3$ and $m=-1,n=3$ are circled.}\label{fig2}
 \end{figure}

\begin{figure}
 \center\includegraphics[width=25pc]{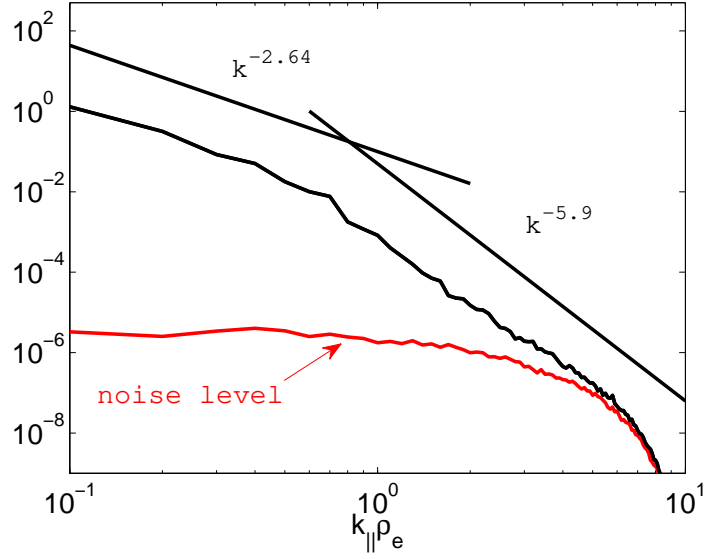}
 \caption{Spectrum of magnetic fluctuations $|\delta B|^2/|B_0|^2$ in the parallel direction $k\rho_{e\parallel}$. The noise level curve is in red. The power law best fits are superimposed.}\label{fig3}
 \end{figure}

\begin{figure}
 \center\includegraphics[width=25pc]{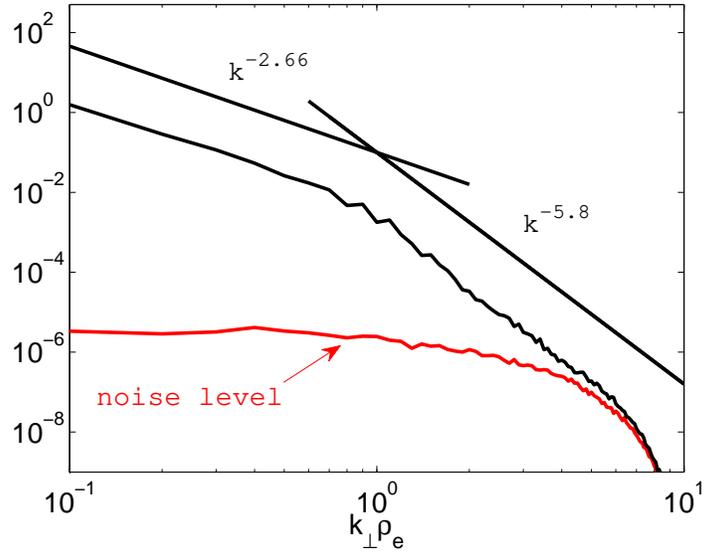}
 \caption{Spectrum of magnetic fluctuations $|\delta B|^2/|B_0|^2$ in the perpendicular direction $k\rho_{e\perp}$. The noise level curve is in red. The power law best fits are superimposed.}\label{fig4}
 \end{figure}

\begin{figure}
 \center\includegraphics[width=25pc]{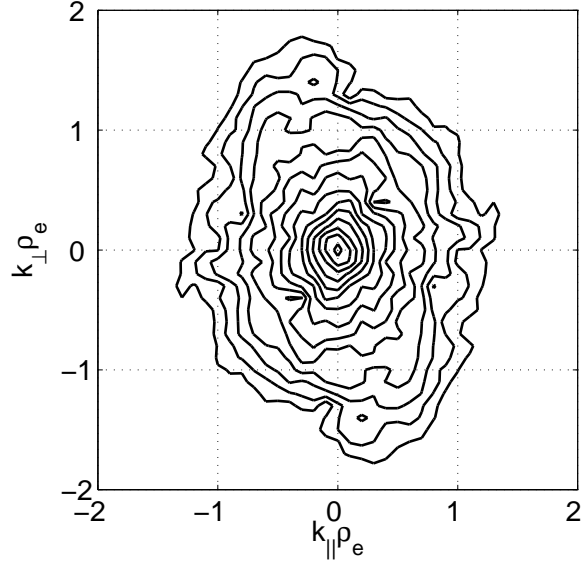}
 \caption{Contour plot of the spectrum of magnetic fluctuations $|\delta B|^2/|B_0|^2$ in parallel and perpendicular directions. The contours are for values of $10^{-15},10^{-14},\ldots,10^{0}$}\label{fig5}
 \end{figure}

The first results that we show are for benchmark cases, where we have initialised the perturbation ($\delta B/B_0\sim 1$) with a single wavenumber and not with a spectrum of waves. Figure \ref{fig1} shows the developed spectrum of magnetic fluctuations $\delta B/B_0$ in wavenumber space, at the end of the simulation. The only mode initially perturbed was the mode $m=1,n=3$, which is indicated with an arrow. The result indicates that the harmonics of the initial mode get predominantly excited according to a 3-wave resonant interaction ($k=k_1+k_2$). Clearly one cannot speak of turbulent cascade in this case, but the purpose of this simulation is to check whether the code is able to treat in a consistent and satisfactory manner injections of energy at small scales. An issue with PIC codes is indeed the unavoidable presence of noise at small scales due to the discreteness of computational particles. This noise could result in the spurious effect of an inverse cascade of energy from large wavenumbers which is completely numerically generated. One can see that this effect is not important in Figure \ref{fig1}, where the modes at large $k$ that get excited are only the higher harmonics of the initial mode. The price to pay in order to avoid spurious effects at small scales is the use of an unusually large number of particles (6400 particles per cell).\\
A confirmation of the fact that 3-wave resonant interactions are well described in the code is shown in Figure \ref{fig2}. In this case we have excited, along with the mode $m=1,n=3$, its counter-propagating mode $m=-1,n=3$. The modes initially excited are circled. The resulting distribution of energy satisfies again the relation $k=k_1+k_2$, giving rise to that checkerboard plot.\\
As we have said, these results should only be considered as benchmark runs. However, from Figure \ref{fig1} and \ref{fig2} it is already clear that the treatment of small amplitudes fluctuations at large $k$ as linear modes poses some problems when the system allows the presence of finite amplitude perturbations at small $k$. This is because nonlinear interactions that couple small and large $k$ modes could always take place, resulting in a continuous injection of energy at small scales. Hence, in order to characterise the dissipation range solely through some linear damping mechanism, one needs to find out at what scale the nonlinear coupling becomes completely ineffective. In other words, the fact that the amplitude of fluctuations is small at some scales does not justify alone the validity of linear theory, in the solar wind. Another requisite is that those scales should not take part in any nonlinear coupling with larger scales.\\
We now show and comment on the results of a run where a flat spectrum of modes has been excited, accordingly to Eqs. (\ref{1}-\ref{3}). The range of modes excited is $(0,3)$ in both wavenumbers $m$ and $n$. As we said, we have tried different slopes for the initial spectrum, but we will comment on one run only, as the results were very similar.\\
Figure \ref{fig3} and \ref{fig4} show the spectra of $|\delta B|^2/|B_0|^2$ in parallel and perpendicular directions (i.e. integrated along the other direction), respectively. The red curve indicates the noise level, which is the level of fluctuations generated by particle noise, when no initial perturbation is imposed in the equilibrium. As it is clear, results above $k\rho_e\sim7$ are not reliable, because the curves approach the noise level. Moreover, in order to be far enough from the noise level, and allow the cascade to proceed as much as possible, we initialise the perturbation with a relatively large amplitude ($\delta B/B_0\sim3$), at $k\rho_e=0.1$. This is not completely consistent with solar wind observations, but it does not affect the purpose of this study, because the amplitude of fluctuations above $k\rho_e=1$ is small enough, in principle, to justify the comparison with linear theory which is presented in the next section. The spectra presented in Figures \ref{fig3} and \ref{fig4} are fully converged. The cascade proceeds very quickly and reaches a stationary profile after about $T\Omega_e\sim 30$. 
Of course, in reality, there may be evolution on time scales
longer, and length scales larger than those simulated. This
could only be investigated with considerably larger and
longer simulations.\\ 
Two interesting features emerge from Figure \ref{fig3} and \ref{fig4}. First, the spectrum develops a well defined power law that steepens at about the inverse electron gyroradius. The plots show that the least-squared fit with a power law function yields exponents which are consistent with solar wind observations (in particular $|\delta B/B_0|^2\sim k^{-2.6}$ between the ion and electron inverse gyroradii, although this feature might be coincidental due to the 2D character of the simulations). Second, the nonlinear cascade proceeds far into the electrons scales, until it reaches the noise level. There is no sign of an exponential roll-over or the emergence of linear dissipation mechanisms, even when the level of $\delta B/B_0$ is as low as $10^{-2}$, which usually considered small enough for applying linear theory.\\ 
The nonlinear cascade proceeds as it is expected in magnetized plasma \citep{horbury08}. It develops the well-known power anisotropy, which means that most of the energy resides, at the end of the simulation, in oblique or quasi-perpendicular modes. This is visible in Figure \ref{fig5}, where we show the spectrum of magnetic perturbations as a contour plot in wavevector space.
\subsection{Comparison with linear theory}
\begin{figure}
 \center\includegraphics[width=25pc]{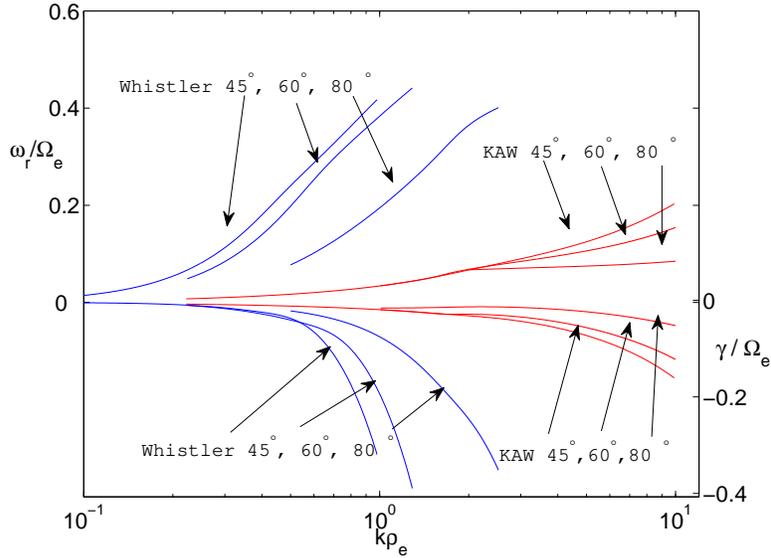}
 \caption{Dispersion curves for KAW and whistler for different angles of propagation: $\theta=45^\circ, 60^\circ, 80^\circ$. Frequencies are normalized on the electron cyclotron frequency $\Omega_{e}$ ($\omega_r$ is the real frequency and $\gamma$ is the damping rate).}\label{fig6}
 \end{figure}

\begin{figure}
 \center\includegraphics[width=25pc]{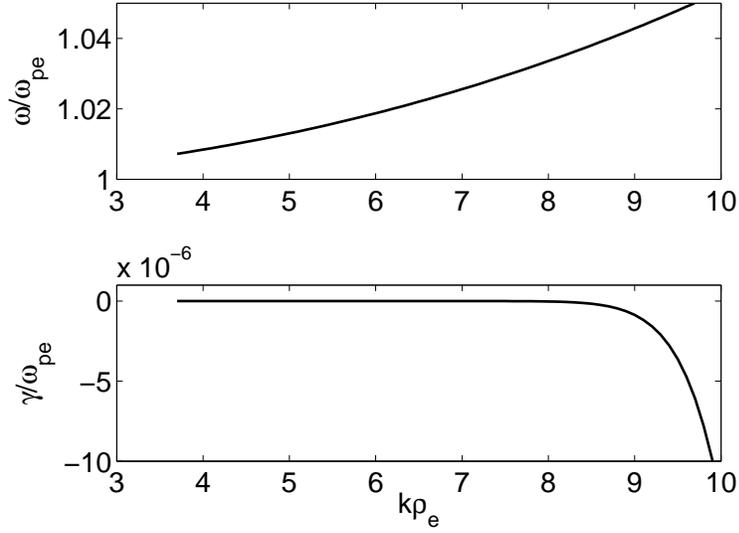}
 \caption{Frequency (top) and damping rate (bottom) for the Langmuir mode at $\theta=80^\circ$, normalized on the electron plasma frequency $\omega_{pe}$.}\label{fig7}
 \end{figure}

In previous works it has been argued that the dynamics of the dissipation at small scales can be described by linear mechanisms, since the amplitude of fluctuations become so small that non-linear effects can be neglected. It is also argued that a purely linear damping mechanism will result in an exponential roll-over of the spectrum of turbulent fluctuations.
In this section, we attempt to interpret the simulation results at the light of Vlasov linear theory. We have used a Vlasov solver to identify linear modes in the range of wavevectors of interest for this study. We anticipate that, at these scales, it becomes very difficult to follow the numerical solution of the dispersion relation for one single mode, as many branches of the solutions overlap. We show in Figures \ref{fig6}-\ref{fig7} the dispersion relations for whistler modes, KAW and Langmuir waves, for different angle of propagation ($\theta=45^\circ,60^\circ,80^\circ$). We focus attention on oblique modes, since we know that the cascade tends to develop the aforementioned anisotropy. We point out that what we call Langmuir waves are the generalization of electrostatic Langmuir modes in presence of magnetic field, studied in \citet{willes00}. They are lightly damped high-frequency oscillations, which are not usually taken in account in this regime.\\
The same dispersion curves for the damping rate only of KAW and whistler are plotted in log-log scale in Figure \ref{fig8}. \citet{li01} have proposed an empirical fit for the damping rate $\gamma$ of KAW of the form:
\begin{equation}\label{fit}
 \gamma=-m_1 k^{m_2}e^{-\left(\frac{m_3}{k}\right)^2},
\end{equation}
where $m_1,m_2,m_3$ are positive fitting parameters. They have also argued, on the basis of a simplified model, which treats the nonlinear cascade as an isotropic diffusion process in wavenumber space, that in order to recover a power-law for the spectrum of magnetic fluctuations, the damping rate should be itself a power-law as a function of $k$. What is interesting, in Figure \ref{fig8}, is that at this scale the damping rate seems to follow precisely a power law in $k$. The form of Equation \ref{fit} could still be valid, but the exponential factor becomes essentially very close to 1, as $k$ becomes very large.\\
Another interesting feature of Figure \ref{fig8}, which has not been previously commented on elsewhere, is that the damping rate of KAW and whistler undergo a smooth transition from a certain power law to another. In principle, one could argue that the dispersion curves of Figure \ref{fig6} and \ref{fig8} might account for the results of Figures \ref{fig3} and \ref{fig4}, i.e. for the cascade of the power spectrum and its steepening, since the properties of the linear waves, at this scale, seem to be favourable for the support of a power law solution for the fluctuations. However, if this is the case, we still need a firm characterisation of the fluctuations in terms of linear waves. \\
Following \citet{gary09} we now try to characterize the linear waves by their electron compressibility $\delta n_e/\delta B$. We prefer to use the electron compressibility rather than the magnetic compressibility, because the former gives a better signature of the difference between KAW and whistler modes. Indeed, for KAWs the electron compressibility grows as a function of $k$, while it decreases for whistlers. In Figure \ref{fig9} we show the values of the electron compressibility calculated from the simulation for different angle of propagation (diamonds for $\theta=45^\circ$, stars for $\theta=60^\circ$, and circles for $\theta=80^\circ$). The solid lines indicates the value predicted for KAWs, whistler and Langmuir modes. The simulations results clearly indicates a trend of increasing values for larger values of $k\rho_e$, which suggests a certain degree of Alfvenicity of the fluctuations. However, none of the linear modes taken in consideration is in good match with the simulation result. The fluctuations are probably a mixture of different modes. Interestingly, the Langmuir mode seem to be the one whose electron compressibility is closest to the simulations results. This might be related to the high level of initial fluctuations. 
\subsection{Particle heating}
The energy at the beginning of the simulations is partitioned equally between the magnetic field and the particles, and electrons and ions have the same kinetic energy. The simulation conserves more than $95\%$ of the total energy, and at the end of the run there is a $5\%$ increment of the total kinetic energy, at the expenses of magnetic energy. The increase in electric energy is negligible. The heating is directed mainly to electrons, such that at the end of the simulation they are $20\%$ hotter than the protons. This is however not surprising, given the time and length scales of the simulations. \\
Electrons are heated predominantly in the parallel direction (with respect to the magnetic field). The value of temperature anisotropy is not constant over the whole computational domain, but depends on the spatial position, and $T_\parallel/T_\perp$ ranges from 0.98 to 4.9. This suggest that the heating is connected with the formation of magnetic topological structures. An interesting feature that emerges by examining the electron distribution function is the presence of non-thermal features. However, the deviation from a bi-Maxwellian distribution is more accentuated when the particles and the mean magnetic field are considered in a small portion of the box. We show in Figure (\ref{pdf_xy}) the electron distribution function in the $(x,y)$ plane for four different nested boxes. Velocities are normalized to the speed of light. The solid line indicates the direction of the mean magnetic field within each box, and $L$ is the length of the boxes, in ion gyroradius units. It turns out that as the magnetic field and the particles are averaged over larger and larger boxes, the distribution function tends to become more and more bi-Maxwellian and any non-thermal feature is lost. A similar result holds in the $(x,z)$ plane, while the distributions in the direction orthogonal to the mean magnetic field remains roughly gyrotropic. This result has important implications for satellite observations. Indeed, it suggests that the time window over which the particles and the magnetic field are averaged affects the distribution functions and therefore the anisotropy measured. By assuming a steady wind flowing at 250 km/s, one can transform the box lengths into frequencies at which data are collected, and estimate what is the level of anisotropy measured at different frequencies. The upper bound of such anisotropies are shown, for electrons, in Figure (\ref{anisotropy}). Given the small length of the computational box used in this study, the range of frequencies plotted in Figure (\ref{anisotropy}) is relatively high. However, future missions and instruments may be able to collect data at such frequencies, and therefore will be able to verify this prediction.

\section{Conclusion}
\begin{figure}
 \center\includegraphics[width=25pc]{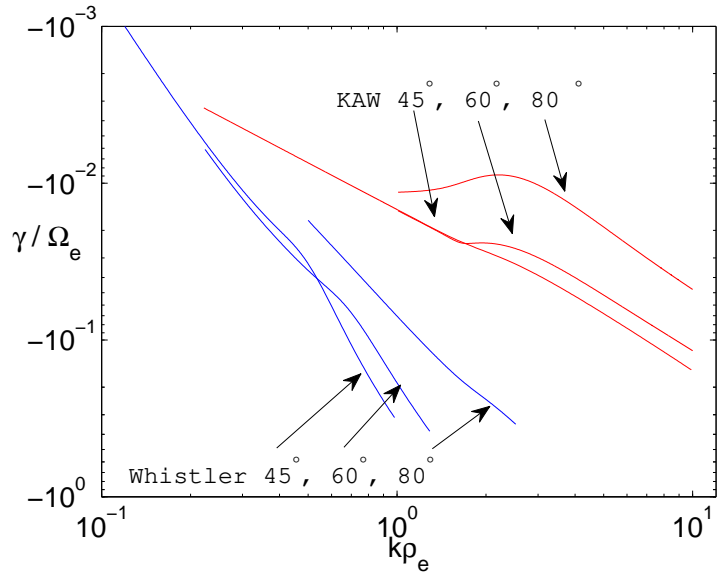}
 \caption{Damping rates for KAW and whistler for different angles of propagation: $\theta=45^\circ, 60^\circ, 80^\circ$, in log-log scale, normalized on the electron cyclotron frequency $\Omega_e$.}\label{fig8}
 \end{figure}

\begin{figure}
 \center\includegraphics[width=25pc]{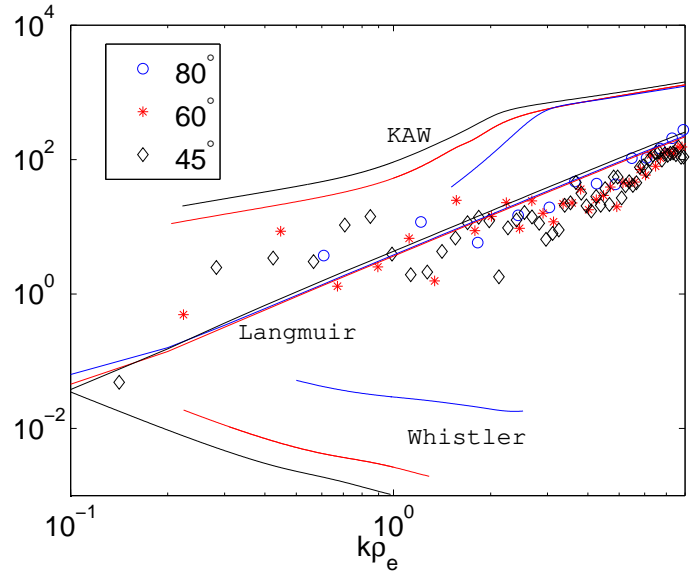}
 \caption{(Color online) Electron compressibility $|\delta n_e|/|\delta B|$. The symbols are the results of simulations (circle for $\theta=80^\circ$, star for $\theta=60^\circ$, diamond for $\theta=80^\circ$). Solid lines are the predictions from linear Vlasov theory for KAW, whistler and Langmuir waves (in blue for $\theta=80^\circ$, red for $\theta=60^\circ$, and black for $\theta=80^\circ$).}\label{fig9}
 \end{figure}

\begin{figure}
 \center\includegraphics[width=25pc]{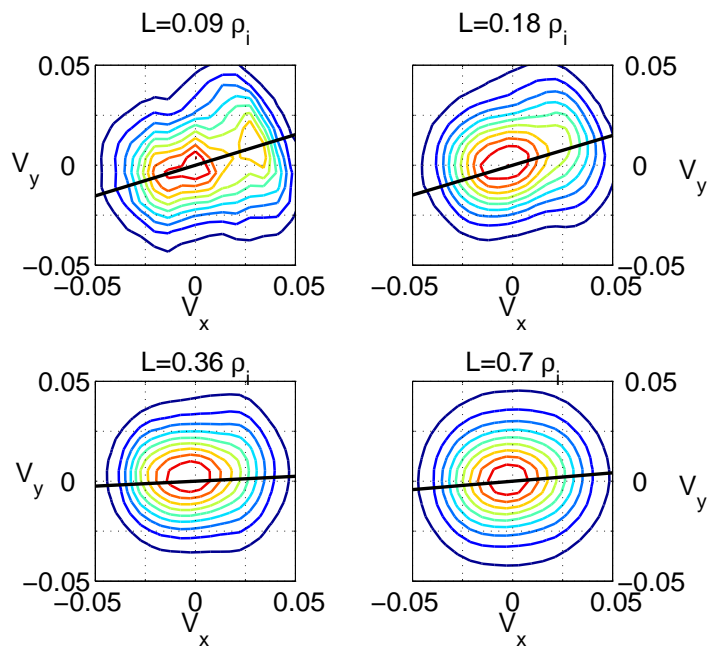}
 \caption{(Color online) Electron distribution functions in the $(x,y)$ plane, collected in four nested boxes of increasing size $L$. The solid line shows the direction of the mean magnetic field within each box. Velocities are normalized to the speed of light.}\label{pdf_xy}
 \end{figure}

\begin{figure}
 \center\includegraphics[width=25pc]{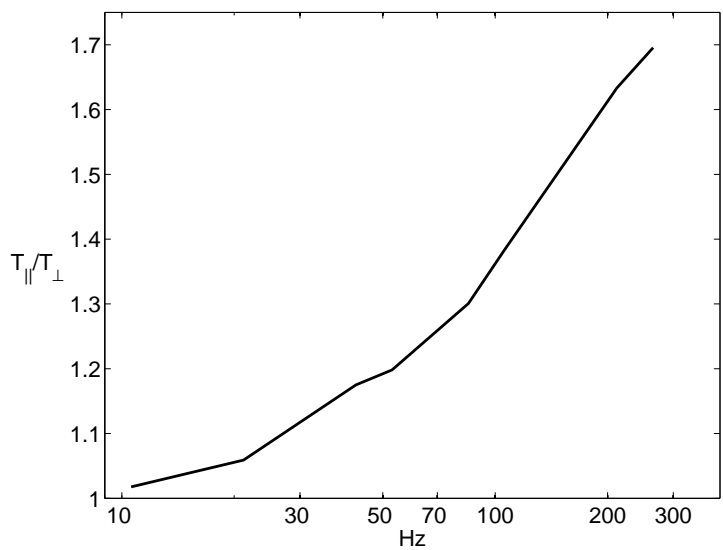}
 \caption{Electron anisotropy $T_\parallel/T_\perp$ versus the frequency at which data are collected, in Hz.}\label{anisotropy}
 \end{figure}

We have performed two-dimensional fully-kinetic PIC simulations of decaying magnetic fluctuations at electron scales. The scope of this work is to address some open questions regarding the dissipation of turbulent fluctuations at small scales, and the plausibility of treating such fluctuations in the framework of linear theory.\\
We have shown the results of the first simulations that use realistic parameters in a fairly large computational domain that extends to electron scales. This represents an improvement in the context of the previous 2D simulations.\\
The results of our simulations suggest that a turbulent cascade in the form of a power-law could proceed very far below the electron gyroradius. It is interesting that at such scales the damping rate for both kinetic Alfven waves and whistler modes are power laws as a function of $k$. In principle, this would suggest that purely linear mechanisms could be responsible for a power law spectrum of fluctuations, such as the one observed in satellite data, and obtained in the PIC simulations. However, we have shown that neither KAW or whistler have an electron compressibility which is in agreement with simulations results.\\
Our interpretation is that many different modes should be taken in account at this scale, including Langmuir waves and possibly other more heavily damped modes. Unfortunately, the level of excitation of each mode depends on non-linear mechanisms active at larger scale. Therefore we doubt that the dissipation process can be understood and described solely within a linear description. However, the puzzling steepening of the power spectrum might be caused by changes of the linear properties of the waves, as the ones reported in Figure \ref{fig8}.\\
Concerning the heating of the particles, we have observed a preferential heating of electrons along the parallel direction (the same happens, in smaller measure, to protons, but the size of the box does not really allow to draw any informative conclusions on protons). It has also emerged that non-thermal features in the electron distribution function appear when the box over which quantities are averaged is not too large (i.e. smaller than 0.2-0.3 $\rho_i$). On the basis of this result we have made a prediction of the upper bound value of electron anisotropy related to the frequency at which data are collected, which can be verified by future measurements.\\
A last comment on the computational aspects of this work is due. The results presented in this paper have been obtained essentially thanks to the use of a semi-implicit code. This has allowed to use a computational box which comprehends scales as little as $k\rho_e=10$ and as large as $k\rho_i\sim4$. However, in order to keep a low level of noise we had to use a large number of particles per cell. 
At present, this excludes the possibility of extending the box so to fully include ion scales, and constrains the geometry to two-dimensional. 
We used a 2D computational spatial domain which contains
the background magnetic field; the particle velocity
space is 3D. This restricts the available $k-$space for
fluctuations to a $(k_\parallel ,k_\perp)$ plane, with the
consequence, particularly, of restricting, but not entirely
eliminating the interaction of modes with $k$ strictly perpendicular to the magnetic field.
However, the use of this particular choice of simulation
plane retains fluctuations with a $k_\parallel$ component
and the possibility of particle parallel
scattering and heating, and is thus a more realistic
choice than having the background magnetic field
strictly perpendicular to the simulation plane. For
these reasons, we believe the 2D results we have presented
are qualitatively robust, but, of course, they will only
be confirmed by fully three-dimensional simulations.
Of course, this is now at the edge of computational feasibility, but a possibility which is worth investigating in the future is to perform an implicit $\delta f-$PIC simulation, so that the number of particles needed could be reduced.\\
 
%
%
\acknowledgments
This work was supported by STFC grant ST/H002731/1.

%
%
%
%
%
%
%
%



%
%
%

\newpage
\begin{table}
\small
\begin{tabular}{||c|c|c|c|c|c|c||}\hline
 Acronym & Reference &  $k\rho_i$ & $\omega_i/\Omega_i$ & $m_i/m_e$ & $N_p$ & Type\\ \hline
  --     & This work &  $4.28 - 428.48$ & 1650 & 1836 & 6400& 2D-3V PIC\\   \hline
  S08    & \citet{saito08} & $0.83-425$ & 96 & 1836 & 64 & 2D-3V PIC\\ \hline
  H08    & \citet{howes08b} & $0.4-8.4$ & - & 1836 & - & 3D-2V Gyrokinetic\\ \hline
  M10    & \citet{markovskii10} &$0.0095-1.21$& 192.3 & - & 1000 & 2D Hybrid\\ \hline
  S09    & \citet{svidzinski09} & $0.03-66.7$ & 15& 100& $>$100 & 2D-3V PIC\\ \hline 
  V10    & \citet{valentini10} & $0.078-10.003$& - & 100 & - & 2D-3V Hybrid-Vlasov\\ \hline
  P09    & \citet{parashar09} & $0.139-35.7$ & - & 25 & 100 & 2D Hybrid\\ \hline
\end{tabular}
\caption{Comparison of different parameters with previous works. $\rho_i$ is the ion gyroradius, $\omega_i/\Omega_i$ is the ratio of ion plasma to cyclotron frequency (a typical value for the solar wind at 1 AU is around 4000), $m_i/m_e$ is the ratio of ion to electron mass, $N_p$ is the number of particles per cell. }\label{tab1}
\end{table}
\normalsize

\end{document}